\documentclass[aps,prl,twocolumn,superscriptaddress,showpacs]{revtex4}
\usepackage{latexsym}
\usepackage{times}
\usepackage{graphicx}
\usepackage{url}
\usepackage[tight,FIGTOPCAP]{subfigure}

\newcommand{\be}{\begin{equation}}
\newcommand{\ee}{\end{equation}} 
\newcommand{\ba}{\begin{eqnarray}}
\newcommand{\ea}{\end{eqnarray}}

\begin{document}
\title{Hydrodynamic induced deformation and orientation of a microscopic
  elastic filament}

\author{M. Cosentino Lagomarsino}
\affiliation{UMR 168 / Institut Curie, 26 rue d'Ulm 75005 Paris, France}
\email[e-mail address: ]{mcl@curie.fr}

\author{I. Pagonabarraga}
\affiliation{Departament de F\'{\i}sica Fonamental, Universitat de
Barcelona, C. Mart\'{\i} i Franqu\'es 1, 08028 Barcelona, Spain}
\email[ e-mail address: ]{ipagonabarraga@ub.edu}
\author{C.P. Lowe}
\affiliation{van 't Hoff Institute for Molecular Science, University of
Amsterdam, Nieuwe Achtergracht 188, 1018 WV Amsterdam, The
Netherlands}
\email[ e-mail address: ]{lowe@science.uva.nl}


\date{\today}

\begin{abstract}
  We describe simulations of a microscopic elastic filament immersed in a
  fluid and subject to a uniform external force.  Our method accounts for the
  hydrodynamic coupling between the flow generated by the filament and the
  friction force it experiences.  While models that neglect this coupling
  predict a drift in a straight configuration, our findings are very
  different.  Notably, a force with a component perpendicular to the filament
  axis induces bending and perpendicular alignment. Moreover, with increasing
  force we observe four shape regimes, ranging from slight distortion to a
  state of tumbling motion that lacks a steady state. We also identify the
  appearance of marginally stable structures. Both the instability of these
  shapes and the observed alignment can be explained by the combined action of
  induced bending and non-local hydrodynamic interactions.  Most of these
  effects should be experimentally relevant for stiff micro-filaments, such as
  microtubules.
\end{abstract}
\maketitle

Semi-flexible polymers and filaments are important components of biological
systems.  All cytoskeletal filaments, used by cells for transport, morphology
and force generation, fall into this category~\cite{How01}. So do their
assemblies, cilia and flagella, used to generate
propulsion~\cite{Grey1955,Lowe2003}. Recent developments in experimental
techniques allow the controlled synthesis, manipulation and direct
visualization of both real~\cite{Goldstein,chu,Stracke2002} and model
filaments~\cite{Gast2004} of this type.  Consequently, there is renewed
interest in the theory and simulation of both their static and dynamic
properties, the structure and dynamics of DNA in flows~\cite{chu} for example.
If one is interested in dynamics, as we are here, one must consider that these
filaments are normally suspended in a fluid.  On the micron scale, inertia is
generally negligible; any motion takes place in an environment effectively a
billion times stickier than we experience in our daily lives~\cite{Purcell}.
The dynamic response is determined by both elastic and fluid forces.
Analytically, this is a non-trivial nonlinear problem.  Nonetheless, at the
linearized level the case of an elastic filament waved at one end has been
solved~\cite{Goldstein}, and the predictions compared successfully with
micro-manipulation experiments on an actin filament.  The fluid was accounted
for by introducing friction coefficients for parallel and perpendicular
filament motion, independent of both the location along the filament and its
configuration. Such an approximation is termed resistive force
theory~\cite{Grey1955} and is satisfactory for several problems. Given
experimentally observed flagella waveforms it gives good predictions for the
swimming speed of spermatozoa~\cite{Lowe2003}. It also predicts the buckling
instability~\cite{Shelley2001} observed in sheared suspensions of filaments.

However, resistive force theory is an approximation.  In reality, a moving
filament sets up spatially varying flow fields in the fluid that couple back
to the motion of the filament itself. To see why this coupling need not be
trivial, suppose we approximate a filament of length $L$ as a set of $n$
beads separated by a fixed distance $b=L/(n-1)$, in the spirit of the ``shish
kebab'' model~\cite{Doi} of a cylinder.
A bead moving with velocity ${\bf v}$ experiences a
hydrodynamic frictional force ${\bf F}_{\rm H} = -\gamma_0 ({\bf v} - {\bf
v}_{\rm H})$, where $\gamma_0$ is a bead friction coefficient and ${\bf
v}_{\rm H}$ is the velocity of the fluid at the location of the bead.
Writing ${\bf v}_{\rm H}$ in terms of
the flow fields of the other beads gives the
hydrodynamic force on bead $i$
\begin{equation}
  \label{fhydro}
  {\bf F}_{i{\rm H}} = -\gamma_0 {\bf v}_i +
  \frac{ \gamma_0}{8 \pi \eta} \sum_{j \neq i } \left[ \frac{{\bf
        F}_j}{r_{ij}} + 
    {\bf F}_j \cdot \frac{ {\bf r}_{ij} {\bf r}_{ij} }{ {r_{ij}}^3 }
  \right] \,
\end{equation}
where ${\bf r}_{ij}$ is the vector connecting beads $i$ and $j$, ${\bf F}_j$
the non-hydrodynamic force acting on bead $j$ and $\eta$ the viscosity of the
fluid~\cite{oseen}.  Consider now a filament subject to a uniform external
force density $\tilde{F}^{\rm x}$ directed perpendicular to the axis of the rod (all forces
and velocities are in the plane defined by the rod and the external force).
If $b << L$ equation~(\ref{fhydro}) gives
\begin{equation}
\label{fhydro2}
F_{{\rm H}}(s) =  -\gamma_0 \left[ v(s) - \frac{ \tilde{F}^{\rm x}}{8 \pi \eta}
 \ln{\frac{s(L-s)}{b^2}} \right]
\end{equation}
where $s \in [0,L]$ is the distance along the filament. The second term in the
bracket is due to the non-local flow created by the rest of the beads. For a
rigid body, the velocity is a constant, while the hydrodynamic friction force
is greater towards the ends of the rod than in the middle. Thus, one expects a
flexible filament to bend. In this letter, we examine this problem numerically
and show that this spatially varying interaction between the filament and the
fluid substantially enriches the dynamic behavior of the system.  Notably,
the simulations predict hydrodynamic induced distortion and orientation that
should be observable experimentally.

We describe results from a simple numerical model that takes into
account the interaction of the filament with its surrounding fluid more
realistically than the resistive force approximation.  The position-dependent
frictional forces are calculated explicitly from the flow fields generated by
the filament and will generally be configuration dependent and non-uniform.
The semiflexible filament is treated as a curve, with its instantaneous
configuration specified by a position vector ${\bf r}(s)$. The length is
fixed and the elastic energy characterized by a Hamiltonian $H=\kappa \int_0^L
C(s)^2 ds/2$~\cite{Christopher}, where $C(s)=\vert \partial^2{\bf r}/\partial
s^2 \vert$ is the local curvature and $\kappa$ the stiffness.  Numerically,
the filament is modeled as a set of rigidly connected beads.  The force
acting on a bead is the sum of the external, tension, bending and hydrodynamic
force. We calculate the latter from eq.~(\ref{fhydro}), with ${\bf F}_j$ being
the sum of all non-hydrodynamic forces acting on bead $j$.
We can use eq.~(\ref{fhydro2}) to calculate the friction
coefficients of a rigid filament modeled to this degree of approximation,
yielding the friction coefficient $\gamma^\perp$ for motion perpendicular to
the rod $\gamma^\perp={4 \pi \eta L}/{\ln{\left( L/b \right)}}$, in the limit
$b<<L$~\cite{gammaperp}.  If the bead spacing is interpreted as the cylinder
radius this result is the same as the exact slender body hydrodynamic theory
for a cylinder~\cite{limitation}.  An advantage of our approach is that,
despite its simplicity, with increasing number of beads it approaches the
correct result for a slender body, including the non-local nature of the
hydrodynamic forces.  It is accurate for very slender filaments and most of
the biofilaments we are interested in satisfy this condition.  Although a
mathematically more sophisticated approach to treat the slender limit
exists~\cite{shelley2}, it does not account for the fact that the hydrodynamic
force for a filament under a uniform field need not be uniform. As the force
on a particle due its hydrodynamic interaction with its neighbors depends
only on the external force and the instantaneous configuration of the
filament, the simple implicit method described in ref.~\cite{Lowe2003}
suffices to integrate the equations of motion.  These are solved subject to
the condition that filament inertial
effects are irrelevant (the motion is overdamped, consistent with neglecting
the fluid inertia).

\begin{figure}
  \includegraphics[scale=0.3,angle=0]{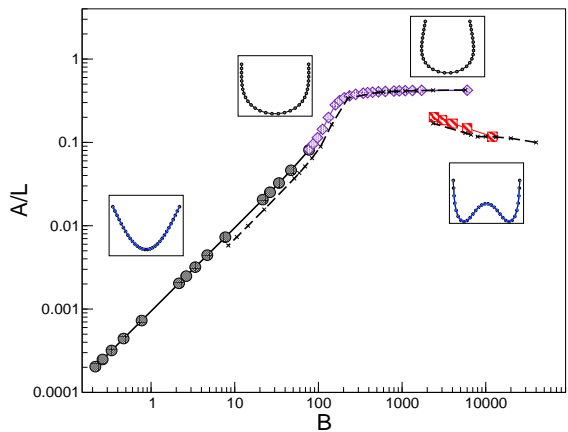}

  \caption{Bending amplitude $A$, as a function of the
    the dimensionless force $B$, for a filament with $L/b=30$ (solid line) and
    $L/b=100$ (dashed line). The insets are
    the corresponding shapes (the external force acts in the downward direction).
    \label{fig:ampl}}
\end{figure}
\begin{figure}
  \includegraphics[scale=0.3,angle=0]{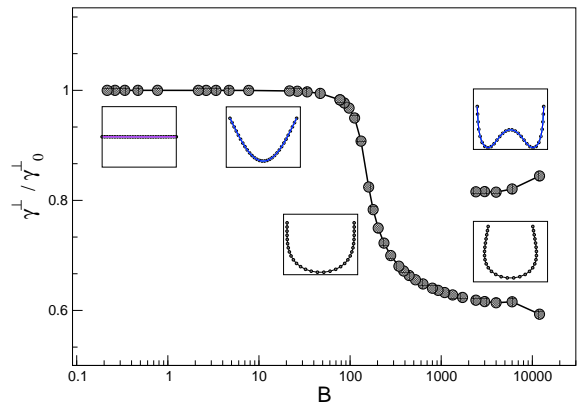}

  \caption{The perpendicular friction coefficient, $\gamma^\perp$, relative to
    its value in the stiff limit $\gamma^\perp_0$, as a function of the
    dimensionless force $B$.
    \label{fig:fric}}
\end{figure}

We first consider the motion of a filament under the action of a uniform field
$\tilde{F}^{\rm x}$, acting perpendicular to the axis
of an initially straight
filament.  The filament's shape evolves in time until a steady state is
reached, where it drifts at a constant velocity with a fixed shape.  That is,
the bending, external and tension forces are balanced by the configuration
dependent fluid force at every point along the filament.  This steady state is
a function of a dimensionless force $B=L^3 \tilde{F}^{\rm x}/\kappa$.  When
$B<<1$, we expect the filament to behave as a rigid rod. With increasing $B$,
significant bending will be required to balance any non-uniformity in the
hydrodynamic force.  Typical shapes we observe for the steady state over a
range of values of $B$ are shown inset in Fig.~\ref{fig:ampl}.  In the steady
state the filament is bent, indicating that the higher frictional force
towards the end of the filament has to be balanced by a bending force.  In the
figure we plot a characteristic transverse distortion, $A$, defined as the
distance between the uppermost and lowermost point of the filament along the
direction of the applied force.  We can identify four distinct regimes.  For
forces corresponding to small values of $B$ $(B < 50)$, the degree of
distortion is small and increasing in proportion to $B$. This is the linear
regime, where the bending can be understood as one of the lowest elastic
modes~\cite{Goldstein,check} excited in response to the applied force (coming
both from the external field and hydrodynamic interactions).  For $B > 100$
the coupling between hydrodynamic and elastic forces is clearly non-linear, as
evidenced by the rounded ``U'' shape of the filament in this regime.  The
bending amplitude saturates and, with increasing $B$, the filament ``U'' shape
becomes increasingly rounded.  In this regime we also see that the friction
coefficient starts to markedly decrease, as shown in Fig.~\ref{fig:fric}.
This is because the progressive alignment of the filament with the applied
force leads to an increasing fraction of parallel motion (see
Fig.~\ref{fig:ampl}), characterized by a lower friction. The perpendicular
friction coefficient is always greater than the parallel friction coefficient
(not shown) but, for large $B$, appears to approach it~\cite{parallel}. Thus,
in this limit the filament approaches hydrodynamically isotropic behavior.
For even higher values of $B$, $(B > 2000$), we see yet different behavior.
The filament initially adopts a new ``W'' shape, that, incidentally, has a
more uniform hydrodynamic force density (and hence is closer to resistive
force theory) and a lower distortion (shown by the lower data points in the
figure) than its final state.  The W configuration is marginally stable; for
any initial perturbation, after a transient time the W configuration rotates
with the same handedness as the original perturbation and finally adopts a
highly distorted but stable ``horse shoe'' shape (the upper data points shown
in Fig.~\ref{fig:ampl}).  At still higher $B$ ($> 4000$), we observe a regime
(not shown) where following the formation of the ``W'' the filament again
rotates but never adopts a new stable state.  Rather, it exhibits a periodic
zig-zagging motion indicating that above some critical value of $B$ there is
no dynamically stable steady state that the filament can reach~\cite{movies}.

\begin{figure}
  \includegraphics[scale=0.3,angle=0]{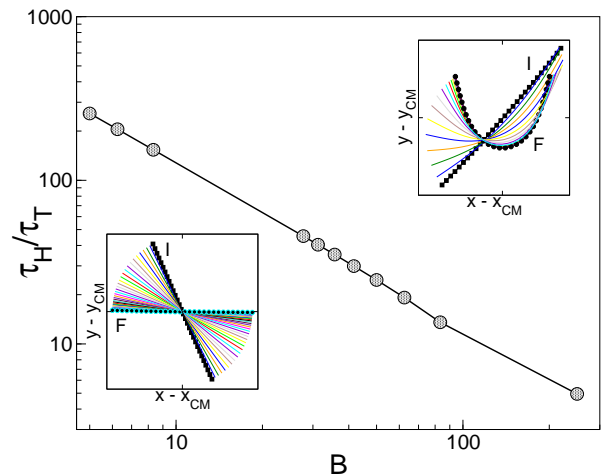}

  \caption{ Log-log
    plot of the characteristic hydrodynamic re-orientation time relative to
    the translational time, $\tau_{\rm H}$, as a function of the dimensionless
    force $B$. Inset are the corresponding motions for two characteristic
    values of B, viewed in the center of mass frame of reference, the initial
    and final states being $I$ and $F$ respectively (the external force acts in the downward direction).
    \label{fig:time}}
\end{figure}

Now we consider a filament that is initially straight but with its axis tilted
with respect to the force.  If the hydrodynamic friction along the rod is
uniform, the filament will remain straight and maintain the same orientation.
It will move at the constant velocity for
which the friction and external forces balance each other~\cite{dhont}.
Because the parallel and perpendicular friction coefficient differ, this will
generally be at an angle to the external force. A truly rigid rod, even with
non-local hydrodynamic interactions, must also move at constant speed
maintaining its initial orientation, as is the case for any rigid object.
However, as shown in Fig.~\ref{fig:time}, we find that if the initial
configuration is rotated anti-clockwise with respect to the external
force, as the
rod translates it rotates clockwise until its ends are again aligned
perpendicular to the force. During the rotation the filament exhibits a
transient drift motion in a direction inclined to the force but in the steady
state, unlike a rigid rod, the drift is along the applied force direction. We
recover identically the case discussed above.

The orientation angle as a  function of time - after a
time during which bending is established - decays exponentially~\cite{thesis}.
The characteristic time for this decay, $\tau_{\rm H}$, relative to the time a
rigid rod takes to translate its length ($\tau_{\rm T}=
\gamma^{\perp}/\tilde{F}^{\rm x}$) is plotted as a function of $B$ in
Fig.~\ref{fig:time}.

This effect is related to the flexibility, because a truly rigid rod cannot
display this behavior.  Why should a flexible filament behave so differently?
As long as the rod is not aligned parallel to the applied force, there is a
component of the force perpendicular to the filament inducing bending.  Taking
the situation illustrated in Fig.~\ref{fig:time}, the bending will slightly
align the tangent to the filament at the right hand end parallel to the
applied force and the tangent at the left hand end perpendicular. The local
friction coefficient is higher perpendicular to the filament than parallel.
So, even if the two ends move with the same velocity, the drag force on the
right hand end will be slightly lower that that on the left. Thus, a torque
tends to rotate the filament clockwise, as we observe (see
Fig.~\ref{fig:time}).  A simple semi-quantitative analysis indicates that in
the linear regime, where the degree of bending is proportional to $B$, the
torque generated is proportional to $B$. One therefore expects that the time
it takes the filament to reorient scales as $1/B$, as shown in
Fig.~\ref{fig:time}.  Note that this simple argument only invokes the
non-local hydrodynamics to generate the bending that breaks the symmetry
between the two ends.  Arguing along the same lines, resistive force level
hydrodynamics predicts rotation of a bent filament.  It fails because it does
not include the mechanism that generates the bending.

This analysis has some interesting consequences.  It appears that {\em any}
degree of elasticity will induce the rotation of the filament. Hence, the
completely rigid limit is singular, in the sense that the absence of
orientation is because the reorientation time becomes indefinitely long as the
rigidity increases. It also implies that the motion of a flexible filament
aligned parallel to the force will be unstable. A small perturbation of the
angle from the parallel, with a certain handedness, generates a torque with
the same handedness. The perturbation will be amplified and the filament will
rotate until the dynamically stable perpendicularly aligned ``U'' state is
reached.  This reasoning also explains why the ``W'' shape is
marginally stable. By symmetry, the two outer ``U'' sections cannot
contribute any net torque. However, the central convex section will display an
instability with respect to any perturbation from the perpendicular,
consistent with our observation that the ``W'' configuration eventually
rotates to form a horse shoe.

Could this behavior be relevant in practice?  Appreciable bending requires
$B>>1$ while hydrodynamic induced rotation should occur for any value of $B$
(although for small $B$ the reorientation time will be long).  In the case of
biofilaments, a gravitational field only gives low values of $B$.
Microtubules have a stiffness of $\kappa \sim 50 pN \mu
m^2$~\cite{Jans_Dog04}, so for a typical length ($10 \mu m$) we estimate that
$B<10^{-3}$.  Accelerations three orders of magnitude greater than gravity are
needed to reach even $B \sim 1$ (which can be achieved in a centrifuge).
However, bio-filaments are commonly charged, and hence will react to applied
electric fields.  If we concentrate on microtubules, they are negatively
charged with an effective charge density $\tilde{q}$ of approximately $200 e /
\mu m$~\cite{Stracke2002}, where $e$ is the elementary charge.  The strongest
electric fields for which microtubules are stable, reported in
ref~\cite{Stracke2002}, are of order $10^4 V/m$, corresponding to a force
density of $0.3 pN/\mu m$.  For a microtubule of length $L\sim 5 \mu m$ (as
used in the experiments reported in~\cite{Stracke2002}) one then finds $B \sim
1$. However, since $B \propto L^3$ for longer microtubules of $L\sim 30 \mu
m$, we have $B \sim 200$.  It is therefore possible to apply forces that
should induce observable bending.  A further condition to observe the
orientational behavior we describe is that the hydrodynamic orientational time
$\tau_{\rm H}$ is shorter than any other relaxation time in the problem.  For
example, we have thus far ignored rotational diffusion. It will tend to
randomize the orientation on a characteristic time scale $\tau_D \sim
\gamma^{\perp} L^2/k T$. From Fig.~\ref{fig:time} we see that $\tau_{\rm H} \sim
\gamma^{\perp}/(\tilde{F}^{\rm x} B)$, so $\tau_{\rm H} / \tau_D \sim L/ (B^2
\lambda$), where $\lambda (= \kappa / k T)$ is the persistence length.  One
can therefore neglect diffusion if $L/(B^2 \lambda) << 1$.  For a stiff
filament, $L/\lambda << 1$ (for a $10 \mu m$ microtubule this ratio is
$10^{-2}$), so this condition is automatically satisfied if $B \geq 1$.  In
addition, microtubules have an electric dipole.  This will lead to a torque
tending to align them parallel to the electric field while they translate and
bend. The ratio of the dipole re-orientation time $\tau_{\rm d}$ to the
translational time is $\tau_{\rm d}/\tau_{\rm T} = \tilde{q} L^2/d$, where $d$
is the electric dipole.  Experimental results indicate that $d \sim e L$
~\cite{Stracke2002}. Hence the ratio is $\tau_{\rm d}/\tau_{\rm T} \sim 200
L/\lambda_{\rm d}$ with the length $\lambda_{\rm d} = 1 \mu m$. So for a
microtubule with $L > 1 \mu m$ the dipolar re-orientation should be negligible
compared to hydrodynamic orientation.

An analysis along these lines suggests that for actin and DNA it is possible
to achieve $B >> 1$. However, the condition that the hydrodynamic orientation
time is shorter than all other time-scales is not easily satisfied. There will
be a competition between the effects we describe here, acting to distort and
align the filament, and thermal effects, acting to randomize the orientation
and maintain the equilibrium structure.  Finally, the situation with carbon
nanotubes is more flexible, given the greater control of physical properties
of these objects~\cite{nanotubes}.  It may be possible to reach the $B>>1$
regime where the instability sets in.

In this letter we have shown how the non-local nature of the hydrodynamic
interactions affects the dynamics of inextensible elastic filaments subject to
a uniform external field.  The method captures all the relevant
configurational couplings. Although the hydrodynamic treatment is only exact
for an infinitely slender rod, it is computationally simple compared to other
techniques~\cite{shelley2,sd}.  The non-uniform, configuration-dependent
friction the fluid exerts on the filament induces distortion and a
corresponding increase in mobility.  In the dynamic steady state, the degree
of bending depends on the stiffness of the filament, a fact that could be used
experimentally to determine $\kappa$.  There is a crossover from the linear
regime to a plateau value for the degree of distortion, reminiscent of other
pattern-forming systems. Our numerical results suggest that at still higher
degrees of forcing there exist long-lived marginally stable states and that
eventually the filament behaves in an unsteady but hydrodynamically isotropic
manner.  The fact that fluid friction induces bending means that a flexible
rod aligns perpendicular to an applied force.  This is not expected if one
neglects either the flexibility of the filament or the non-local
hydrodynamics. It does, however, offer a novel route by which one may
manipulate the orientations of filaments experimentally.

\begin{acknowledgments}
  This work is part of the research program of the ``Stichting voor
  Fundamenteel Onderzoek der Materie (FOM)'', which is financially supported
  by the ``Nederlandse organisatie voor Wetenschappelijk Onderzoek (NWO)''.
  I.P. acknowledges financial support from DGICYT of the Spanish Government
  and DURSI of the Generalitat de Catalunya and thanks the FOM Institute for
  its hospitality. C.L. thanks E.~Koopman for help with the movies.
\end{acknowledgments}


\begin{thebibliography}{999}
\bibitem{oseen} By using the Oseen tensor we disregard the beads' size and
  consider them hydrodynamically as points of force.
\bibitem{gammaperp} The friction coefficient $\gamma^\perp$ is defined as the
  ratio of the total external force acting on a straight filament to an
  imposed uniform drift velocity. Using the same definition, the parallel
  friction coefficient is $\gamma^\perp/2$, again consistent with slender body
  theory.
\bibitem{limitation} Our model cannot represent cylindrical filaments
with a given aspect ratio using a variable number of beads. This
requires an explicit filament/fluid interface which is computationally
much more demanding.
\bibitem{check} We verified that for small amplitudes the elastic modes
  can be projected onto the known elastic eigenmodes of the filament.
\bibitem{parallel} The parallel friction coefficient is computed from a
  straight configuration parallel to the field, before any
  instability occurs.
\end{thebibliography}
\end{document}